\documentclass{article}

\usepackage{amsmath}
\usepackage{amssymb}
\usepackage{amsfonts}
\usepackage{bm}
\usepackage[dvips,xdvi]{graphicx}

\newcommand{\bs}[1]{\boldsymbol{\mathsf{#1}}}
\newcommand{\real}{\operatorname{\mathrm{Re}}}
\newcommand{\imag}{\operatorname{\mathrm{Im}}}

\newlength{\refwidth}
\begin{document}


\title{\bf Normal Modes of Rouse-Ham Symmetric Star Polymer Model}
\author{Takashi Uneyama \\
Department of Materials Physics, Graduate School
 of Engineering, 
Nagoya University, \\
Furo-cho, Chikusa, Nagoya 464-8603, Japan}


\maketitle

\begin{abstract}
The Rouse-Ham model is a simple yet useful dynamics model for an unentangled branched polymer. In this work, we study the normal modes of the Rouse-Ham type coarse-grained symmetric star polymer model. We model a star polymer by connecting multiple arm beads to a center bead by harmonic springs. In the Rouse-Ham model, the dynamics of the bead positions can be decomposed into the normal modes, which are chosen to be orthogonal to each other. Due to the existence of degenerate eigenvalues, the eigenmodes do not directly correspond to the normal modes. We propose several methods to construct the normal modes for the coarse-grained symmetric star polymer model. We show that we can construct the normal modes by using a simple permutation or the Hadamard matrix. These methods give symple and highly symmetric orthogonal modes, but work just for a special number of arms. We also show that we can construct the normal modes by using the discrete Fourier transform (DFT) matrix. This method is applicable for an arbitrary number of arms.
\end{abstract}


%

\section{INTRODUCTION}

Coarse-grained simple dynamics models are sometimes quite useful to
study various dynamical and rheological behaviors of polymeric materials.
One simple yet useful model is the Rouse model\cite{Rouse-1953,Doi-Edwards-book},
in which a linear polymer chain is modeled by connecting beads by linear springs.
Beads do not interact each other except via the harmonic spring potentials,
and they obey the Langevin equation with a constant friction coefficient.
The Rouse model can be generalized to polymers with more complex architectures
such as star polymers and comb polymers. Ham\cite{Ham-1957} developed
such a generalized model and how the chain architecture affects the
linear viscoelasticity.

The Rouse-Ham model is a standard model to 
analyze the dynamics of an unentangled branched polymer.
For example, the viscoelastic and dielectric relaxation behaviors of
unentangled star chains can be well described by the Rouse-Ham model\cite{Watanabe-Yoshida-Kotaka-1990}.
Even for entangled polymers, some long-time dynamical behaviors
can be well described by the Rouse-Ham model governed by
the constraint release (CR) mechanism\cite{Watanabe-1999,Watanabe-Matsumiya-Inoue-2002}.
(Strictly speaking, the CR-based Rouse model is different from the
Rouse model based on the Langevin equation. At the long-time scale,
nonetheless, they exhibit almost the same relaxation behavior.\cite{Uneyama-2024})
Recently, Zhang and coworkers\cite{Zhang-Tang-Chen-Kwon-Matsumiya-Watanabe-2024a,Zhang-Tang-Chen-Kwon-Matsumiya-Watanabe-2024b}
utilized the Rouse-Ham model to study nonlinear rheological behaviors of
an end-associative tetra-arm star polymer solution.

We have one difficulty when we utilize the
Rouse-Ham model to analyse star polymer dynamics.
If the star polymer is symmetric (if all the arms have the
same molecular weight), some eigenmodes of the Rouse-Ham model are degenerate.
This is in contrast to the Rouse model for a linear chain, where all the
eigenmodes are not degenerate.
If we are interested only on linear viscoelasticity, this degeneracy is
not serious. We need only the eigenvalue distribution to calculate the
linear viscoelasticity. However, if we want to study the chain dynamics
explicitly, care is required. This is because the degenerate eigenmodes
are not orthogonal in general. We will be required to construct orthogonal normal modes by combining non-orthogonal eigenmodes.

In this work, we consider how to construct orthogonal normal modes
for the Rouse-Ham type symmetric star polymer model. We propose three different methods to systematically construct
orthogonal eigenmodes.
The first method is based on the permutation of one specific eigenvector.
This method works only for a tetra-arm star polymer.
The second method is based on the Hadamard matrix\cite{Zwillinger-book}.
This works for other numbers of arms, but the applicability is still limited.
The third method is based on the discrete Fourier transform (DFT) matrix\cite{Rao-Yip-book,Grady-Polimeni-book}.
Unlike other two methods, this works for any number of arms.
After we show the construction methods, we briefly discuss some possible
applications of proposed methods.

\section{MODEL}
\label{model}

We model a symmetric star polymer with $n$-arms by $(n + 1)$ beads connected
by harmonic springs.
We assume that $n \ge 3$.
We describe the position of the $j$-th bead as $\bm{r}_{j}$ ($j = 0 , 1, 2, \dots, n$).
One bead ($j = 0$) represents the branch point, and other beads
($j = 1, 2, \dots, n$) represent free ends of arms.
We may call the zeroth bead as the center bead and
other beads as the arm beads.
The coarse-grained effective interaction energy is 
\begin{equation}
 \label{effective_interaction_energy}
 \mathcal{U}(\lbrace \bm{r}_{j} \rbrace) = \sum_{j = 1}^{n}
  \frac{3 k_{B} T}{2 \bar{Q}^{2}} (\bm{r}_{j} - \bm{r}_{0})^{2}.
\end{equation}
Here, $k_{B}$ is the Boltzmann coefficient, $T$ is the temperature,
and $\bar{Q}$ is the equilibrium root mean square arm size.
In the Rouse-Ham model, the beads obey the (overdamped) Langevin equation
with a constant friction:
\begin{equation}
 \label{langevin_equation}
 \frac{d\bm{r}_{j}(t)}{dt} = - \frac{1}{\zeta} \frac{\partial \mathcal{U}(\lbrace \bm{r}_{j}(t) \rbrace)}{\partial \bm{r}_{j}(t)} + \bm{\xi}_{j}(t),
\end{equation}
where $\zeta$ is the friction coefficient and $\bm{\xi}_{j}(t)$ is the Gaussian
white noise. The noise satisfies the fluctuation-dissipation relation:
$\langle \bm{\xi}_{j}(t) \rangle = 0$, $\langle \bm{\xi}_{j}(t) \bm{\xi}_{k}(t') \rangle = 2 k_{B} T\delta_{jk} \bm{1} \delta(t - t') / \zeta$
($\langle \dots \rangle$ represents the statistical average and $\bm{1}$ is the unit tensor).
The Brownian force (which has the dimension of the force) which appear in the
underdamped Langevin equation corresponds to $\zeta \bm{\xi}_{j}(t)$.
By substituting eq~\eqref{effective_interaction_energy} into eq~\eqref{langevin_equation},
we have
\begin{equation}
 \label{langevin_equation_with_rouse_ham_matrix}
 \frac{d\bm{r}_{j}(t)}{dt} 
= - \frac{3 k_{B} T}{\zeta \bar{Q}^{2}} \sum_{k = 0}^{n} A_{j,k} \bm{r}_{k}(t)
+ \bm{\xi}_{j}(t),
\end{equation}
where $A_{j,k}$ is the $(j,k)$ element of the $(n + 1) \times (n + 1)$ matrix $\bs{A}$ defined as
\begin{equation}
 \label{rouse_ham_matrix}
 \bs{A} =
 \begin{bmatrix}
 n & -1 & -1 & \dots & -1 \\
 -1 & 1 & 0 & \dots & 0 \\
 -1 & 0 & 1 & \dots & 0 \\
 \vdots & \vdots & \vdots & \ddots & \vdots \\
 -1 & 0 & 0 & \dots & 1
\end{bmatrix}.
\end{equation}
In what follows, we call $\bs{A}$ as the Rouse-Ham matrix.
In this work, we use bold sans-serif letters to express vectors and matrices for the bead indices,
in order to distinguish them from the vectors and matrices in the three dimensional space.

Eq~\eqref{langevin_equation_with_rouse_ham_matrix} is linear
in $\lbrace \bm{r}_{j}(t) \rbrace$. Therefore, it is convenient to introduce
the normal modes. With the normal modes, eq~\eqref{langevin_equation_with_rouse_ham_matrix}
can be rewritten as a set of $(n + 1)$ independent Langevin equations.
The normal modes can be constructed by orthogonalizing the Rouse-Ham matrix $\bs{A}$,
in the same way as the Rouse model for a linear chain\cite{Rouse-1953,Doi-Edwards-book}.
To construct normal modes, we calculate the eigenmodes of $\bs{A}$.
We express the $p$-th eigenvector and eigenvalue as $\bs{v}_{p}$ (which is an $(n + 1)$-dimensional column vector) and $\lambda_{p}$ ($p = 0, 1, 2, \dots, n$).
The eigenvector and eigenvalue satisfy $\bs{A} \cdot \bs{v}_{p} = \lambda_{p} \bs{v}_{p}$.
Then the $p$-th eigenmode of eq~\eqref{langevin_equation_with_rouse_ham_matrix}
is calculated as $\bm{X}_{p} \propto \sum_{j = 1}^{n} v_{p,j} \bm{r}_{j}$.
The relaxation time of the $p$-th eigenmode is $\tau_{p} = \zeta \bar{Q}^{2} / 3 k_{B} T \lambda_{p}$
if $\lambda_{p} > 0$.
{\em If all the eigenvectors are orthogonal}, the thus constructed eigenmodes
$\lbrace \bm{X}_{p}(t) \rbrace$ become the orthogonal normal modes.
However, if the Rouse-Ham matrix $\bs{A}$ has degenerate eigenvalues, the eigenvectors
for degenerate eigenvalues are {\em not} orthogonal in general. In such a case,
$\lbrace \bm{X}_{p}(t) \rbrace$ cannot be employed as the normal modes
unless they are properly orthogonalized.

\begin{figure}[tbh]
\begin{center}
 \includegraphics[width=1.08\refwidth]{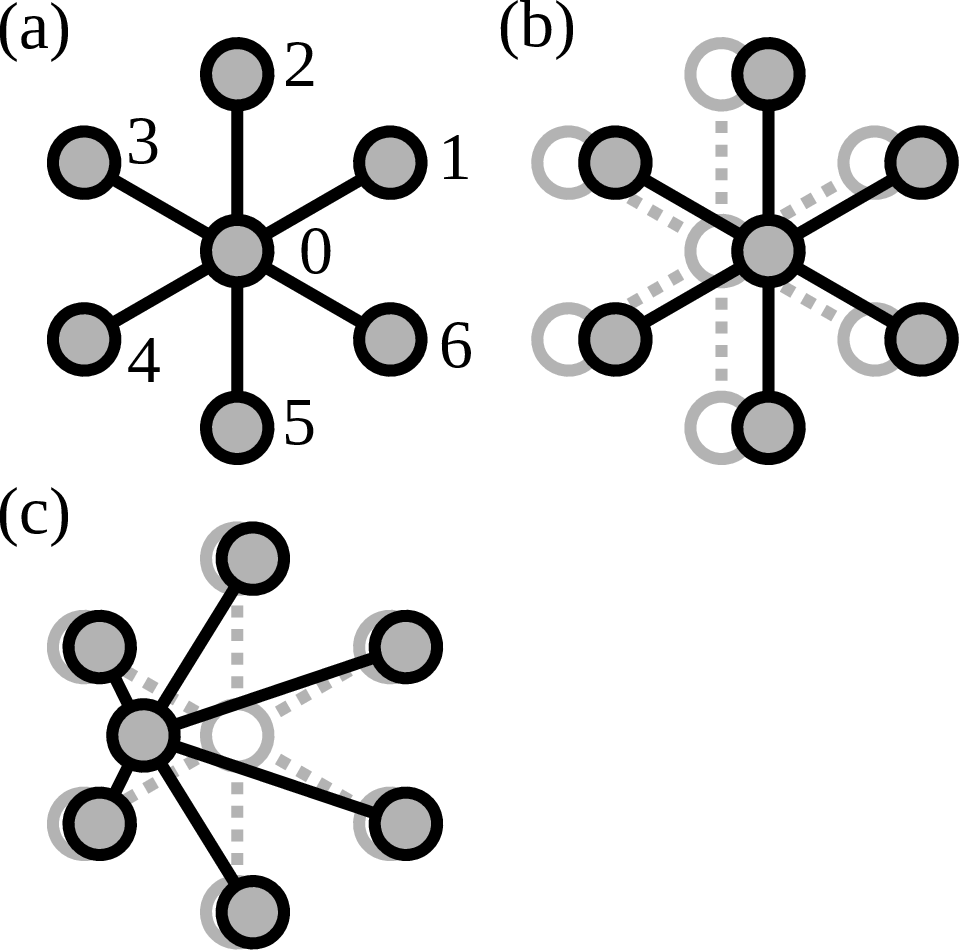}%
\end{center}
\caption{The zeroth and $n$-th normal modes for a symmetric Rouse-Ham star polymer.
 As an example, a hexa-arm star ($n = 6$) is shown here.
 (a) The bead indices and reference
 bead positions. Black circles filled with gray and black lines represent beads and springs, respectively.
 Beads positions modulated by the (b) zeroth and (c) $n$-th normal modes.
 The zeroth mode induces the translation while the $n$-th mode 
 induces the deformation.
 The displacements of beads by normal modes are superposed on the reference
 bead positions. Only the displacements of beads
 in the horizontal direction are considered.
 For comparison, the reference bead positions are shown as
 open gray circles. 
 \label{hexa_star_nondegenerate_modes}}
\end{figure}

The eigenvalues and eigenvectors of $\bs{A}$ given by eq~\eqref{rouse_ham_matrix} can be calculated rather easily,
if we do not care about the orthogonality.
The Rouse model for a linear chain has the zeroth mode with the zero eigenvalue,
which corresponds to the center of mass motion. In the same way,
we have the zeroth eigenvector and eigenvalue as
$\bs{v}_{0} = [1 \ 1 \ 1 \ \dots \ 1]^{\mathrm{T}}$ (the superscript ``T'' represents the transpose) and
$\lambda_{0} = 0$.
Then eq~\eqref{langevin_equation_with_rouse_ham_matrix}
has the zeroth normal mode as the center of mass position:
\begin{equation}
 \label{zeroth_normal_mode}
 \bm{X}_{0}(t) = \frac{1}{n + 1} \sum_{j = 0}^{n} \bm{r}_{j}.
\end{equation}
The remaining $n$ eigenvalues are non-zero.
The eigenvalues for $1 \le p \le n - 1$ are degenerate: $\lambda_{1} = \lambda_{2} = \dots = \lambda_{n - 1} = 1$.
The $n$-th eigenvector and eigenvalue are
$\bs{v}_{n} = [- n \ 1 \ 1 \ \dots \ 1]^{\mathrm{T}}$ and
$\lambda_{n} = n$. The $n$-th normal mode is given as
\begin{equation}
 \label{nth_normal_mode}
 \bm{X}_{n}(t) = \sum_{j = 1}^{n} \bm{r}_{j}(t) - n \bm{r}_{0}(t).
\end{equation}
Fig.~\ref{hexa_star_nondegenerate_modes} shows images for the translation and 
deformation by the zeroth and $n$-th normal modes, respectively.

Since the eigenmodes for $p = 1, 2, \dots, n - 1$ are degenerate, the
explicit forms of eigenvectors
($\bs{v}_{1}, \bs{v}_{2}, \dots, \bs{v}_{n - 1}$) are not that clear.
Ham\cite{Ham-1957} showed that so-called the span vectors correspond to
the eigenmodes.
The $p$-th span vector can be constructed as $\bm{X}_{p}^{(\text{H})}(t) = \bm{r}_{p}(t) - \bm{r}_{p + 1}(t)$
($p = 1,2,\dots,n - 1$).
Unfortunately,  Ham's span vectors are {\em not} orthogonal, and
thus we cannot employ the Ham span vectors as normal modes. The $j$-th element of
the $p$-th eigenvector $\bs{v}_{p}$, which corresponds to the $p$-th
span vector, becomes
\begin{equation}
 \label{span_vector_element}
 v_{p,j} = 
  \begin{cases}
   1 & (j = p), \\
   -1 & (j = p + 1), \\
   0 & (\text{otherwise}).
  \end{cases}
\end{equation}
The zeroth element is always zero. Thus it would be convenient to 
express $\bs{v}_{p} = [0 \ s_{p,1} \ s_{p,2} \ \dots \ s_{p,n}]^{\mathrm{T}}$
and use the $n$-dimensional vector $\bs{s}_{p} = [s_{p,1} \ s_{p,2} \ \dots \ s_{p,n}]^{\mathrm{T}}$ instead of $\bs{v}_{p}$ for $1 \le p \le n - 1$.
It would be intuitive to show $\lbrace \bs{s}_{p} \rbrace$ as an 
$(n - 1) \times n$ matrix:
\begin{equation}
 \begin{bmatrix}
  \bs{s}_{1} & \bs{s}_{2} & \bs{s}_{3} & \dots & \bs{s}_{n - 1}
 \end{bmatrix}
 =
 \begin{bmatrix}
  1 & 0 & 0 & \dots & 0 \\
  -1 & 1 & 0 & \dots & 0 \\
  0 & -1 & 1 & \dots & 0 \\
  \vdots & \vdots & \vdots & \ddots & \vdots \\
  0 & 0 & 0 & \dots & -1
 \end{bmatrix}.
\end{equation}
These eigenvectors are clearly not orthogonal because $\bs{v}_{p} \cdot \bs{v}_{p + 1} = \bs{s}_{p} \cdot \bs{s}_{p + 1} = -1$.
In what follows, we simply call $\bs{s}_{p}$ as the $p$-th span vector, since it
can be directly related to the Ham span vector as $\bm{X}_{p}^{\text{(H)}}(t) = \sum_{j = 1}^{n} s_{p,j} \bm{r}_{j}(t)$.

In the case where only the eigenvalue distribution is important,
the existence of orthogonal eigenvectors is sufficient. We do not need
their explicit forms.
For example, to calculate the linear viscoelasticity, we do not need the
explicit expression of the normal modes.
However, if we want to know the dynamics of bead positions,
we will be required to construct orthogonal normal modes.
In principle, we can construct $(n - 1)$ orthogonal eigenvectors starting from
$(n - 1)$ non-orthogonal yet linearly independent eigenvectors, by using such as the Gram-Schmidt orthogonalization\cite{Gautschi-book} and introduction of some small perturbations.
But if we naively apply the Gram-Schmidt orthogonalization to $\lbrace \bs{s}_{p} \rbrace$,
the resulting eigenvectors become very complicated (especially when $n$ is large).
Handling of perturbations is also very complicated.
In Sec.~3, we show that we can construct orthogonal eigenvectors
in three different methods without such complicated processes.

\begin{figure}[tbh]
\begin{center}
 \includegraphics[width=1.03\refwidth]{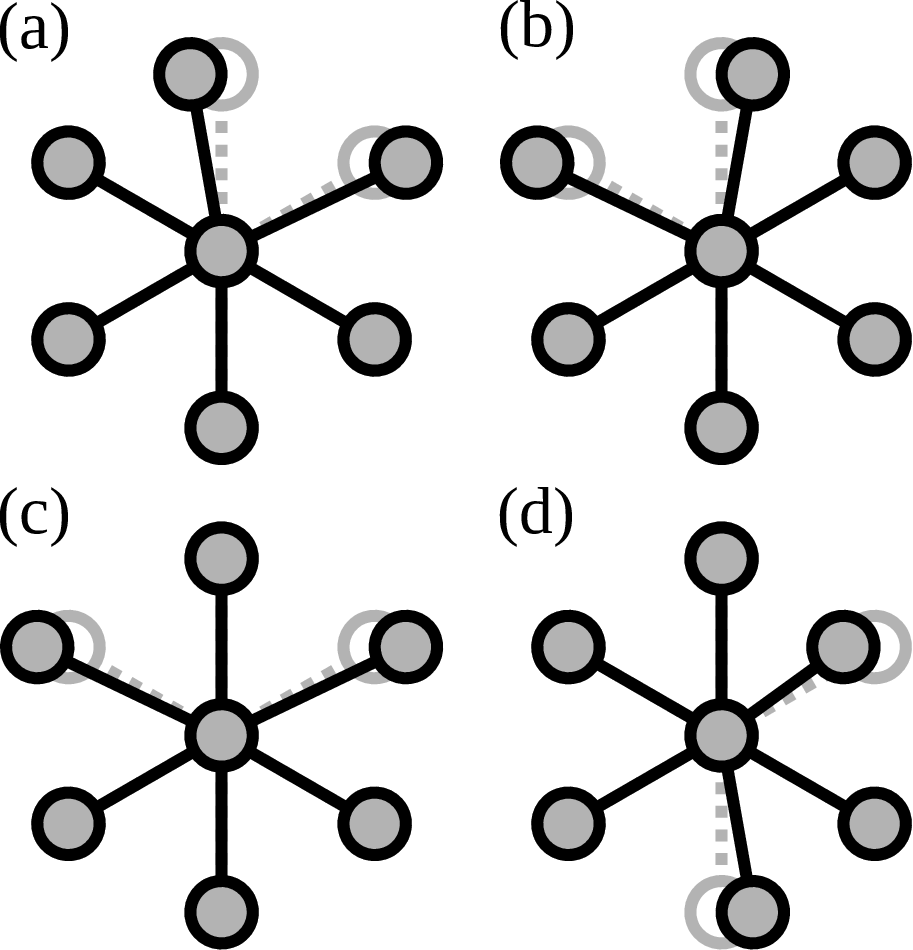}%
\end{center}
\caption{Some degenerate eigenmodes for a symmetric Rouse-Ham star polymer,
 expressed by the span vectors.
 We use a hexa-arm star ($n = 6$) used in 
 Fig.~\ref{hexa_star_nondegenerate_modes}.
 The bead indices and reference bead positions are shown in Fig.~\ref{hexa_star_nondegenerate_modes}(a).
 The span vectors (a) $\bs{s}_{1}$ and (b) $\bs{s}_{2}$.
 (c) A span vector formed by the sum of two span vectors $\bs{s}_{1}$ and $\bs{s}_{2}$
 (eq~\eqref{span_vector_element_shifted} with $p = 1$).
 (d) A span vector in the general form (eq~\eqref{general_span_vector_element} with $p = 5$ and $q = 1$).
 \label{hexa_star_degenerate_modes}}
\end{figure}

Before we proceed to the construction of normal modes, we show some properties of the span vectors $\lbrace \bs{s}_{p} \rbrace$. If we add two
span vectors $\bs{s}_{p}$ and $\bs{s}_{p + 1}$, we have
\begin{equation}
 \label{span_vector_element_shifted}
  s_{p,j} + s_{p + 1,j} = 
  \begin{cases}
   1 & (j = p), \\
   -1 & (j = p + 2), \\
   0 & (\text{otherwise}).
  \end{cases}
\end{equation}
Eq~\eqref{span_vector_element_shifted} corresponds to an eigenvector with the eigenvalue $1$, too, 
and thus it can be used as a new span vector.
By comparing eqs~\eqref{span_vector_element} and \eqref{span_vector_element_shifted},
we find that the position of $-1$ in $\bs{s}_{p}$ is ``shifted'' from $j = p + 1$ to $j = p + 2$.
This is physically natural, because
a symmetric star polymer is symmetric under the exchange of two arms.
In a similar way, we can ``shift'' positions of $1$ and $-1$ several times. Then the general form of the
span vector $\tilde{\bs{s}}$ is
\begin{equation}
 \label{general_span_vector_element}
  \tilde{s}_{j} =
  \begin{cases}
   1 & (j = p), \\
   -1 & (j = q), \\
   0 & (\text{otherwise}),
  \end{cases}
\end{equation}
for $1 \le p \le n - 1$, $1 \le q \le n - 1$, and $p \neq q$.
Fig.~\ref{hexa_star_degenerate_modes} shows some span vectors.

\section{RESULTS}
\label{results}

\subsection{Permutation approach}
\label{permutation_approach}

We consider the case where $n$ is even. We collect $(n / 2)$ general span vectors,
of which non-zero elements are not overlapped each other. Then we can
construct an eigenvector $\bs{s}_{p}^{\text{(perm)}}$, of which element
$s_{p,j}^{\text{(perm)}}$ can take only $1$ or $-1$.
Without loss of generality, we can fix the first element: $s_{p,1}^{\text{(perm)}} = 1$.
Then we can construct other eigenvectors as follows, by considering permutations. We select $(n / 2 - 1)$
elements from $2 \le j \le n$ and set them $1$. We set the remaining
$(n / 2)$ elements to be $-1$.
The number of the eigenvectors constructed in this way is ${}_{n - 1}C_{n / 2 - 1}$.
On the other hand, we should have only $(n - 1)$ orthogonal eigenvectors.
Therefore, the following condition for $n$ should be satisfied: ${}_{n - 1}C_{n / 2 - 1} = n - 1$.
For $n \ge 3$, this condition is satisfied only for $n = 4$.
Therefore, we limit ourselves to the case of $n = 4$.

\begin{figure}[tbh]
\begin{center}
 \includegraphics[width=\refwidth]{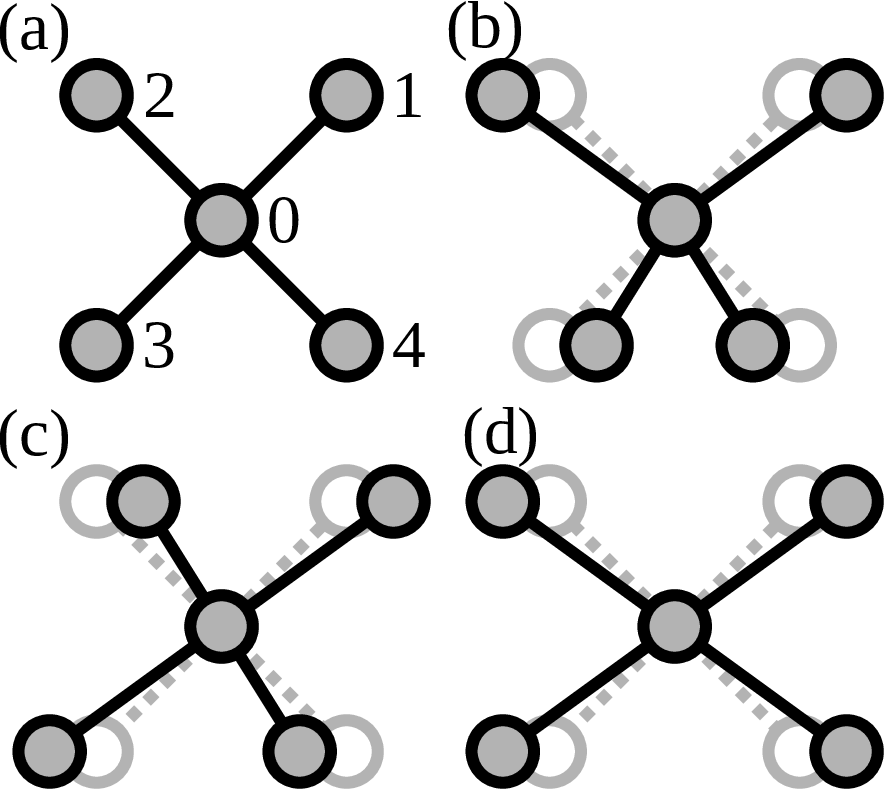}%
\end{center}
\caption{The orthogonal eigenmodes for a tetra-arm symmetric star ($n = 4$),
 calculated by the permutation approach. (a) The bead indices and reference
 bead positions.
 (b)-(d) Deformations by eigenmodes (eqs~\eqref{permutation_eigenvectors_4} and \eqref{normal_mode_from_orthogonal_eigenvectors_perm}) superposed on the reference
 bead positions.  
 \label{tetra_star_normalmodes_permutation}}
\end{figure}

We have the following
permutation-based eigenvectors:
\begin{equation}
 \label{permutation_eigenvectors_4}
 \begin{bmatrix}
  \bs{s}_{1}^{(\text{perm})} &
  \bs{s}_{2}^{(\text{perm})} &
  \bs{s}_{3}^{(\text{perm})}
\end{bmatrix}
=
 \begin{bmatrix}
  1 & 1 & 1 \\
  -1 & 1 & -1 \\
  1 & -1 & -1 \\
  -1 & -1 & 1
\end{bmatrix}.
\end{equation}
It is straightforward to show that these eigenvectors are orthogonal:
$\bs{s}_{p}^{(\text{perm}) \mathrm{T}} \cdot \bs{s}_{q}^{(\text{perm})} = 4 \delta_{pq}$.
Then we can construct the normal modes as
\begin{equation}
 \label{normal_mode_from_orthogonal_eigenvectors_perm}
 \bm{X}_{p}(t) = \sum_{j = 1}^{4} s_{p,j}^{(\text{perm})} \bm{r}_{j}(t) \qquad (p = 1, 2, 3).
\end{equation}
We show images of deformations by orthogonal normal modes by eqs~\eqref{permutation_eigenvectors_4}
and \eqref{normal_mode_from_orthogonal_eigenvectors_perm} in Fig.~\ref{tetra_star_normalmodes_permutation}.

\subsection{Hadamard matrix approach}
\label{hadamard_matrix_approach}

The permutation approach in Sec.~3.1 is simple but not applicable for $n \neq 4$. If we can successfully select $(n - 1)$ eigenvectors which 
are orthogonal to each other from ${}_{n - 1}C_{n / 2 - 1}$ candidates,
we will be able to construct the orthogonal eigenvectors for $n \neq 4$
(of course, $n$ should be even).
Fortunately, for some $n$, there exists such a set of eigenvectors.
We can utilize the Hadamard matrix\cite{Zwillinger-book}.

The Hadamard matrix $\bs{H}_{n}$ is an $n \times n$ matrix of which
elements consist only of $1$ and $-1$. Any two columns (or rows) of $\bs{H}_{n}$
are orthogonal. One column consists only of $1$ (or $-1$). The remaining
($n - 1$) columns can be used as the orthogonal eigenvectors, $\bs{s}^{(\text{H})}_{p}$.
For convenience, we introduce a dummy vector $\bs{s}^{(\text{H})}_{0} = [1 \ 1 \ 1 \ \dots \ 1]^{\mathrm{T}}$.
Then the eigenvectors are given by the following simple form:
\begin{equation}
 \begin{bmatrix}
  \bs{s}_{0}^{(\text{H})} &  \bs{s}_{1}^{(\text{H})} &  \bs{s}_{2}^{(\text{H})} & \dots & \bs{s}_{n - 1}^{(\text{H})}
 \end{bmatrix}
 = \bs{H}_{n}.
\end{equation}

\begin{figure}[tbh]
\begin{center}
 \includegraphics[width=1.15\refwidth]{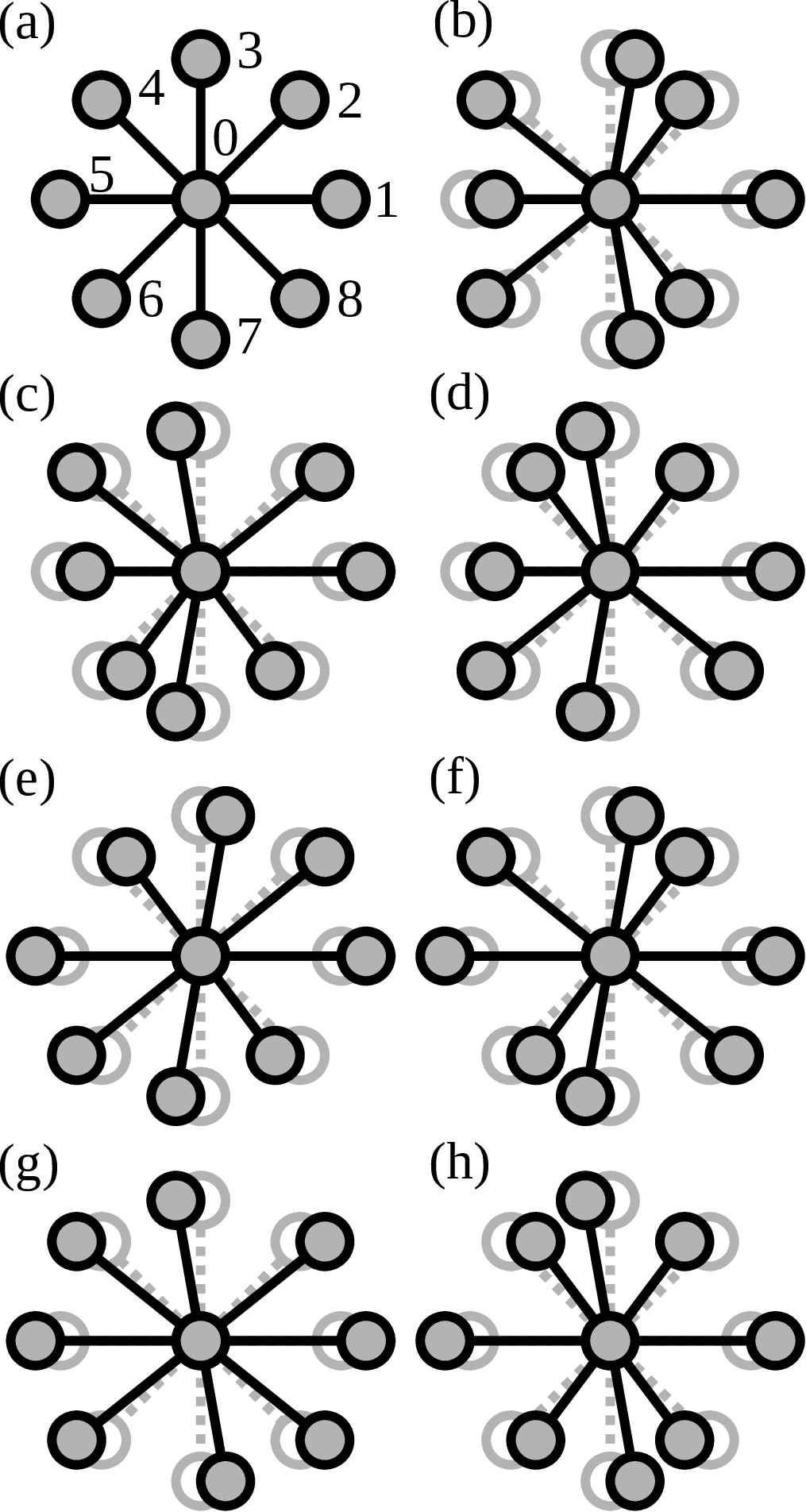}%
\end{center}
\caption{The orthogonal eigenmodes for an octa-arm symmetric star ($n = 8$),
 calculated by the Hadamard matrix approach. (a) The bead indices and reference
 bead positions. 
 (b)-(h) Deformations by eigenmodes superposed on the reference
 bead positions.
 \label{octa_star_normalmodes}}
\end{figure}

Unfortunately, the Hadamard matrix exists only for some limited $n$ ($n = 4, 8, 12, 16, \dots$).
Thus we cannot utilize this construction method for $n = 6$, for example.
We show the explicit forms of the Hadamard matrices for $n = 4$ and $8$:
\begin{equation}
 \label{hadamard_matrix_4}
 \bs{H}_{4} =
  \begin{bmatrix}
   1 &  1 &  1 &  1 \\
   1 & -1 &  1 & -1 \\
   1 &  1 & -1 & -1 \\
   1 & -1 & -1 &  1
  \end{bmatrix},
\end{equation}
\begin{equation}
 \label{hadamard_matrix_8}
 \bs{H}_{8} =
  \begin{bmatrix}
   1 &  1 &  1 &  1 &  1 &  1 &  1 &  1 \\
   1 & -1 &  1 & -1 &  1 & -1 &  1 & -1 \\
   1 &  1 & -1 & -1 &  1 &  1 & -1 & -1 \\
   1 & -1 & -1 &  1 &  1 & -1 & -1 &  1 \\
   1 &  1 &  1 &  1 & -1 & -1 & -1 & -1 \\
   1 & -1 &  1 & -1 & -1 &  1 & -1 &  1 \\
   1 &  1 & -1 & -1 & -1 & -1 &  1 &  1 \\
   1 & -1 & -1 &  1 & -1 &  1 &  1 & -1
  \end{bmatrix}.
\end{equation}
The Hadamard matrices can be numerically constructed by some programs.
(``{\tt hadamard}'' function in Octave constructs and returns the Hadamard matrix\cite{Octave-manual}.
Eqs~\eqref{hadamard_matrix_4} and \eqref{hadamard_matrix_8}
are calculated by Octave 5.2.0.)
The eigenvectors constructed from $\bs{H}_{4}$ are the same
as eq~\eqref{permutation_eigenvectors_4} ($\bs{s}^{(\text{perm})}_{p} = \bs{s}^{(\text{H})}_{p}$).
The normal modes can be constructed in the same way as eq~\eqref{normal_mode_from_orthogonal_eigenvectors_perm}:
\begin{equation}
 \label{normal_mode_from_orthogonal_eigenvectors}
 \bm{X}_{p}(t) = \sum_{j = 1}^{n} s_{p,j}^{(\text{H})} \bm{r}_{j}(t) \qquad
(j = 1, 2, \dots, n - 1).
\end{equation}
Fig.~\ref{octa_star_normalmodes} shows the deformations by the orthogonal normal modes
for an octa-arm star polymer ($n = 8$) by eq~\eqref{hadamard_matrix_8}
and \eqref{normal_mode_from_orthogonal_eigenvectors}.

For $n \ge 8$, we can construct different normal modes by exchanging the
bead indices for the arm beads. The dynamics of beads $\lbrace \bm{r}_{j}(t) \rbrace$
is independent of the choice of the normal modes, and thus any physical quantities
such as the stress tensor are also independent of the choise.

\subsection{DFT matrix approach}

Even if we employ the approach based on the Hadamard matrix
explained in Sec.~3.2,
still we cannot construct the orthogonal eigenvectors in many cases.
For example, we cannot construct the eigenvectors
for $n = 3, 5,$ and $6$. Here we consider another method which is applicable
for any $n \ge 3$, based on the discrete Fourier transform (DFT) matrix\cite{Rao-Yip-book,Grady-Polimeni-book}.

We express the $p$-th orthogonal eigenvector to be constructed as $\bs{s}_{p}^{(\text{F})}$.
The result of Sec.~3.2 implies
that the introduction of a dummy vector which consists only of $1$ is useful.
Thus we introduce it again,
$\bs{s}^{(\text{F})}_{0} = [1 \ 1 \ 1 \ \dots \ 1]^{\mathrm{T}}$,
and construct an $n \times n$ matrix $\bs{S}^{(\text{F})}$ by collecting $n$ eigenvectors:
$\bs{S}^{(\text{F})} = [\bs{s}_{0}^{(\text{F})} \, \bs{s}_{1}^{(\text{F})} \, \bs{s}_{2}^{(\text{F})} \, \dots \, \bs{s}_{n - 1}^{(\text{F})}]$.
Also, we introduce a dummy span vector $\bs{s}_{0} = [-1 \ 0 \ 0 \ \dots \ 1]$, and construct
an $n \times n$ matrix $\bs{S}$ as
$\bs{S} = [\bs{s}_{0} \ \bs{s}_{1} \ \bs{s}_{2} \ \dots \ \bs{s}_{n - 1} ]$.
Its explicit form is
\begin{equation}
 \bs{S} =
  \begin{bmatrix}
   -1 &  1 &  0 &  0 &  \dots &  0 \\
    0 & -1 &  1 &  0 &  \dots &  0 \\
    0 &  0 & -1 &  1 &  \dots &  0 \\
    0 &  0 &  0 & -1 &  \dots &  0 \\
    \vdots & \vdots & \vdots & \vdots & \ddots & \vdots \\
    1 &  0 &  0 &  0 &  \dots &  -1
  \end{bmatrix}.
\end{equation}
The dummy span vector $\bs{s}_{0}$ is {\em not} linearly independent of other span vectors.
Actually, it can be expressed in terms of other span vectors:
\begin{equation}
 \label{dummy_span_vector}
 \bs{s}_{0} = - \sum_{p = 1}^{n - 1} \bs{s}_{p}.
\end{equation}

We consider the eigenvectors of $\bs{S}$.
$\bs{S}$ can be interpreted as the difference operator for a one dimensional
periodic system. We introduce a field on a one dimensional periodic lattice 
with period $n$ as
$f_{j}$. From the periodic condition, we have $f_{j + n} = f_{j}$ and thus
$f_{n} = f_{0}$. We express this field as an $n$-dimensional vector $\bs{f} = [f_{0} \ f_{1} \ f_{2} \ \dots \ f_{n - 1} ]^{\mathrm{T}}$.
If we multiply $\bs{S}$ to $\bs{f}$, we have the difference vector:
\begin{equation}
 \bs{S} \cdot  \bs{f} = 
  \begin{bmatrix}
   f_{1} - f_{0} \\
   f_{2} - f_{1} \\ 
   \vdots \\
   f_{0} - f_{n - 1}
  \end{bmatrix} .
\end{equation}
This means that $\bs{S}$ can be interpreted as a discrete analog of
the differential operator $(d / dx)$ in a continuum one dimensional
periodic system. For a continuum one dimensional periodic system, the eigenmodes of
$(d / dx)$ are the Fourier modes. Even if the system is discrete, the Fourier modes still work
as the eigenmodes.

Therefore, the discrete Fourier modes can be utilized as the orthogonal eigenmodes.
We have the $p$-th eigenvector as
$\bs{s}_{p}^{(\text{F})} = [1 \ \omega_{n}^{p} \ \omega_{n}^{2 p} \ \dots \omega_{n}^{(n - 1) p}]^{\mathrm{T}}$
with $\omega = \exp(2 \pi i / n)$ being the primitive $n$-th root of unity.
$\bs{S}^{(\text{F})}$ becomes the DFT matrix:
\begin{equation}
 \bs{S}^{(\text{F})} = \bs{F}_{n} =
  \begin{bmatrix}
   1 & 1 & 1 & \dots & 1 \\
   1 & \omega_{n} & \omega_{n}^{2} & \dots & \omega_{n}^{(n - 1)} \\
   1 & \omega_{n}^{2} & \omega_{n}^{4} & \dots & \omega_{n}^{2 (n - 1)} \\
   1 & \omega_{n}^{3} & \omega_{n}^{6} & \dots & \omega_{n}^{3 (n - 1)} \\
   \vdots & \vdots & \vdots & \ddots & \vdots \\
   1 & \omega_{n}^{n - 1} & \omega_{n}^{2 (n - 1)} & \dots & \omega_{n}^{(n - 1)^{2}} \\
  \end{bmatrix} .
\end{equation}
It is straightforward to show that $\bs{s}_{p}^{(\text{F})}$ is the eigenvector
of $\bs{S}$ and it satisfies $\bs{S} \cdot \bs{s}_{p}^{(\text{F})} = (1 - \omega_{n}^{p}) \bs{s}_{p}^{(\text{F})}$.
(The columns of a DFT matrix are eigenvectors of a cirulant matrix\cite{Grady-Polimeni-book}.)
For $1 \le p \le n - 1$, $\bs{s}_{p}^{(\text{F})}$ can be constructed as a
linear combination of $\lbrace \bs{s}_{p} \rbrace$:
\begin{equation}
\begin{split}
  \bs{s}_{p}^{(\text{F})}
 & = \frac{1}{1 - \omega_{n}^{p}} \bs{S} \cdot \bs{s}_{p}^{(\text{F})} 
  = \sum_{q = 0}^{n - 1} \frac{\omega_{n}^{pq}}{1 - \omega_{n}^{p}} \bs{s}_{q} \\
 & =  \sum_{q = 1}^{n - 1} \frac{\omega_{n}^{pq} - 1}{1 - \omega_{n}^{p}} \bs{s}_{q} .
\end{split}
\end{equation}
where we utilized eq~\eqref{dummy_span_vector}.
It is also straightforward to show the orthogonality: $\bs{s}_{p}^{(\text{F}) \dagger} \cdot \bs{s}_{q}^{(\text{F})} = n \delta_{pq}$
where $\bs{s}_{p}^{(\text{F}) \dagger}$ represents the Hermitian conjugate vector of $\bs{s}_{p}^{(\text{F})}$.
Thus we find that $\lbrace \bs{s}_{p}^{(\text{F})} \rbrace$
can be utilized as the orthogonal eigenvectors.
The normal modes can be constructed by eq~\eqref{normal_mode_from_orthogonal_eigenvectors}
with $\bs{s}_{p}^{(\text{H})}$ replaced by $\bs{s}_{p}^{(\text{F})}$.

Here, it should be noted that $\bs{s}_{p}^{(\text{F})}$ is a complex vector. It can be
decomposed into the real and imaginary parts. One may suspect that there
are more than $(n - 1)$ eigenvectors if we decompose $(n - 1)$ complex vectors into real vectors.
We show that there are just $(n - 1)$ real eigenvectors.
The complex conjugate of $\bs{s}^{(\text{F})}_{p}$ 
becomes
\begin{equation}
 \label{complex_conjugate_relation_fourier_vector_odd}
 \bs{s}^{(\text{F})*}_{p} = 
  \begin{bmatrix}
   1 \\
   \omega_{n}^{- p} \\
   \omega_{n}^{- 2 p} \\
   \vdots \\
   \omega_{n}^{- (n - 1) p}
  \end{bmatrix}
  =
  \begin{bmatrix}
   1 \\
   \omega_{n}^{(n - p)} \\
   \omega_{n}^{2 (n - p)} \\
   \vdots \\
   \omega_{n}^{(n - 1) (n - p)}
  \end{bmatrix} =
\bs{s}^{(\text{F})}_{n - p}
\end{equation}
Here, $\bs{s}^{(\text{F})*}_{p}$ represents the complex conjugate vector
of $\bs{s}^{(\text{F})}_{p}$, and we utilized the properties of the $n$-th primitive root:
$\omega^{*}_{n} = \exp(- 2 \pi i / n) = \omega^{-1}_{n}$ and
$\omega^{n}_{n} = 1$.
If $n$ is odd, we have
$\real \bs{s}^{(\text{F})}_{p} = \real \bs{s}^{(\text{F})}_{n - p}$
and  $\imag \bs{s}^{(\text{F})}_{p} = - \imag \bs{s}^{(\text{F})}_{n - p}$
($\real$ and $\imag$ represent the real and imaginary parts, respectively)
for $p = 1, 2, \dots, (n - 1) / 2$.
This means that two real eigenvectors constructed from $\bs{s}^{(\text{F})}_{n - p}$ are
exactly the same as those from $\bs{s}^{(\text{F})}_{p}$.
Therefore, we can construct just $(n - 1)$ real eigenvectors from $\lbrace \bs{s}_{p}^{(\text{F})} \rbrace$.
If $n$ is even, we have $ \bs{s}^{(\text{F})*}_{p} =  \bs{s}^{(\text{F})*}_{n - p}$
for $p = 1, 2, \dots,  n / 2 - 1$. Then we can construct $(n - 2)$ real eigenvectors.
For $p = n / 2$, we have
\begin{equation}
 \label{complex_conjugate_relation_fourier_vector_even_center}
 \bs{s}^{(\text{F})}_{n/2} = 
  \begin{bmatrix}
   1 \\
   \omega_{n}^{n/2} \\
   \omega_{n}^{n} \\
   \vdots \\
   \omega_{n}^{- (n - 1) n / 2}
  \end{bmatrix}
  = 
  \begin{bmatrix}
   1 \\
   - 1 \\
   1 \\
   \vdots \\
   -1
  \end{bmatrix},
\end{equation}
where we used $\omega_{n}^{n / 2} = \exp(i \pi) = -1$.
Thus $\bs{s}^{(\text{F})}_{n/2}$ has only the real part ($\imag \bs{s}^{(\text{F})}_{n/2} = 0$), and we can construct just one
real eigenvector from $\bs{s}^{(\text{F})}_{n/2}$.
In total, we can construct $(n - 1)$ real eigenvectors
from $\lbrace \bs{s}_{p}^{(\text{F})} \rbrace$.
Therefore, for any $n$, we have just $(n - 1)$ real eigenvectors.

\begin{figure}[tbh]
\begin{center}
 \includegraphics[width=\refwidth]{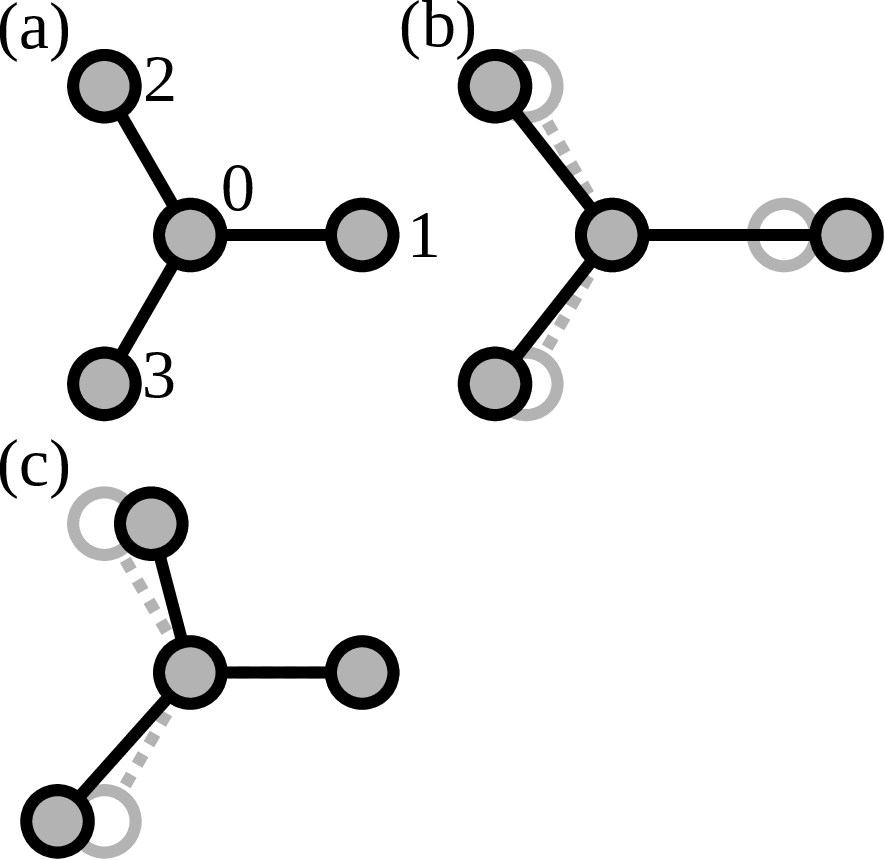}%
\end{center}
\caption{The orthogonal eigenmodes for a tri-arm symmetric star ($n = 3$),
 calculated by the DFT matrix approach. (a) The bead indices and reference
 bead positions. 
 (b), (c) Deformations by eigenmodes superposed on the reference
 bead positions.
 \label{tri_star_normalmodes}}
\end{figure}

\begin{figure}[tbh]
\begin{center}
 \includegraphics[width=0.95\refwidth]{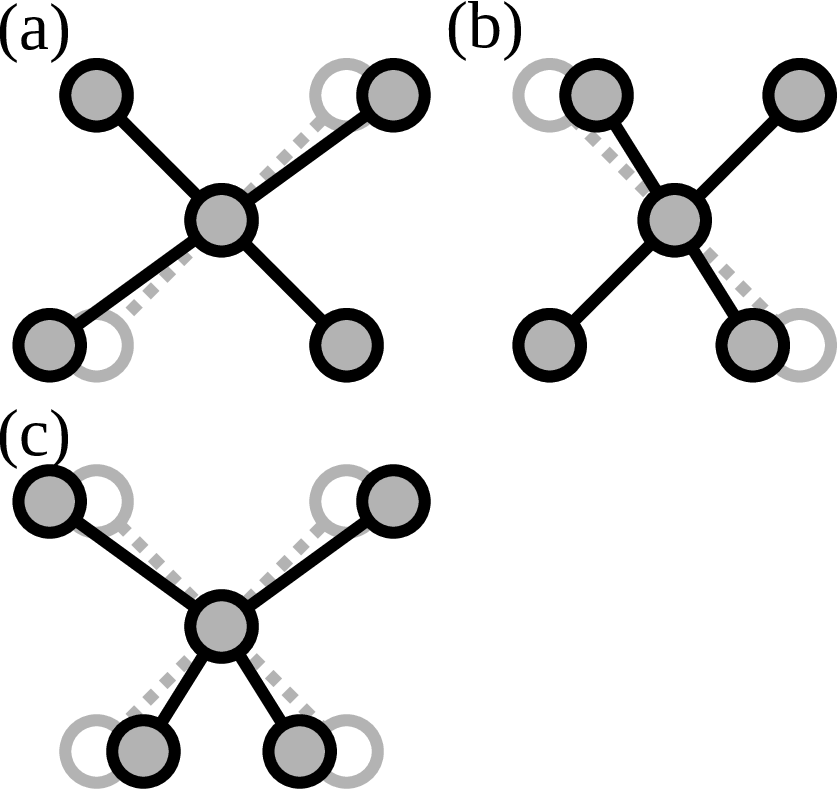}%
\end{center}
\caption{The orthogonal eigenmodes for a tetra-arm symmetric star ($n = 4$),
 calculated by the DFT matrix approach. (a)-(c)
 Deformations by eigenmodes superposed on the reference
 bead positions. The reference bead positions and bead indices are given by Fig.~\ref{tetra_star_normalmodes_permutation}(a).
 \label{tetra_star_normalmodes_fourier}}
\end{figure}

We show some examples below. For $n = 3$, the DFT matrix becomes
\begin{equation}
 \bs{F}_{3}  = 
\begin{bmatrix}
 1 & 1 & 1 \\
 1 & \displaystyle \frac{-1 + \sqrt{3} i }{2} & \displaystyle \frac{-1 - \sqrt{3} i }{2} \\
 1 & \displaystyle \frac{-1 - \sqrt{3} i }{2} & \displaystyle \frac{-1 + \sqrt{3} i }{2} 
\end{bmatrix}.
\end{equation}
The orthogonal real eigenvectors are constructed as
\begin{equation}
 \begin{bmatrix}
  \real \bs{s}_{1}^{(\text{F})} &
  \imag \bs{s}_{1}^{(\text{F})}
 \end{bmatrix}
 =
\begin{bmatrix}
 1 & 0 \\
 -1/2 &  \sqrt{3}/2 \\
 -1/2 & - \sqrt{3}/2
\end{bmatrix}.
\end{equation}
$\imag \bs{s}_{1}^{(\text{F})}$ is essentially the same as the
span vector: $\imag \bs{s}_{1}^{(\text{F})} = (\sqrt{3} / 2) \bs{s}_{2}$.
For $n = 4$, the DFT matrix is given as
\begin{equation}
 \bs{F}_{4}  = 
\begin{bmatrix}
 1 & 1 & 1 & 1 \\
 1 & i & -1 & -i \\
 1 & -1 & 1 & -1 \\
 1 & -i & -1 & i 
\end{bmatrix},
\end{equation}
and we have the following orthogonal real eigenvectors:
\begin{equation}
 \label{fourier_eigenvectors_4}
 \begin{bmatrix}
  \real \bs{s}_{1}^{(\text{F})} &
  \imag \bs{s}_{1}^{(\text{F})} &
  \bs{s}_{2}^{(\text{F})}
 \end{bmatrix}
 =
\begin{bmatrix}
 1 & 0 & 1  \\
 0 & 1 & -1  \\
 -1 & 0 & 1  \\
 0 & -1 & -1 
\end{bmatrix}.
\end{equation}
$\real \bs{s}_{1}^{(\text{F})}$ and $\imag \bs{s}_{1}^{(\text{F})}$ are span vectors, again.
Figs.~\ref{tri_star_normalmodes} and \ref{tetra_star_normalmodes_fourier} show
the deformations by the orthogonal eigenmodes for tri- and tetra-arm star polymers.

The eigenvectors by eq~\eqref{fourier_eigenvectors_4}
do {\em not} coincide to the permutation-based eigenvectors given by eq~\eqref{permutation_eigenvectors_4}.
This is not surprising, because there are many different sets of
orthogonal basis vectors.
Two sets of orthogonal eigenvectors can be related to each other as
\begin{align}
 \bs{s}_{1}^{(\text{perm})} & = \bs{s}_{2}^{(\text{F})}, \\
 \bs{s}_{2}^{(\text{perm})} & =   \real \bs{s}_{1}^{(\text{F})} + \imag \bs{s}_{1}^{(\text{F})}, \\
 \bs{s}_{3}^{(\text{perm})} & = \real \bs{s}_{1}^{(\text{F})} - \imag \bs{s}_{1}^{(\text{F})}.
\end{align}
We can employ both eq~\eqref{permutation_eigenvectors_4} and \eqref{fourier_eigenvectors_4} to construct
normal modes.

In a similar way to the normal modes by the Hadamard matrix, different
normal modes can constructed by exchanging the bead indices for the arm beads.
As before, the physical quantities are independent of the choise of the
indices and normal modes.

\section{DISCUSSIONS}

\subsection{Tetra-arm star chain model with two friction coefficients}

Zhang and coworkers proposed a simple Rouse-Ham type tetra-arm star chain model 
as a coarse-grained model for an end-associative star polymer solution\cite{Zhang-Tang-Chen-Kwon-Matsumiya-Watanabe-2024a,Zhang-Tang-Chen-Kwon-Matsumiya-Watanabe-2024b}.
The chain ends of polymers form transient cross-links,
and a traisient network which exhibits viscoelasticity is formed.
Zhang et al modeled the dynamics of an end-associative star polymer
by tuning the friction coefficient.
In their model, the center bead has much smaller friction coefficient than
the arm beads. As a result, the center bead is equilibrated almost immediately,
and we have the effective dynamic equations only in terms of the arm beads.
In this subsection, we consider their model. We introduce the friction coefficient
\begin{equation}
 \zeta_{j} =
  \begin{cases}
   \zeta / \phi & (j = 0), \\
   \zeta & (1 \le j \le 4),
  \end{cases}
\end{equation}
($\phi \gg 1$) and employ the following Langevin equation instead of eq~\eqref{langevin_equation}:
\begin{equation}
 \label{langevin_equation_two_friction_coefficients}
 \frac{d\bm{r}_{j}(t)}{dt} = - \frac{1}{\zeta_{j}} \frac{\partial \mathcal{U}(\lbrace \bm{r}_{j}(t) \rbrace)}{\partial \bm{r}_{j}(t)} + \bm{\xi}_{j}(t).
\end{equation}
The fluctuation-dissipation relation is
$\langle \bm{\xi}_{j}(t) \rangle = 0$ and $\langle \bm{\xi}_{j}(t) \bm{\xi}_{k}(t') \rangle = 2 k_{B} T\delta_{jk} \bm{1} \delta(t - t') / \zeta_{j}$.
We can rewrite eq~\eqref{langevin_equation_two_friction_coefficients}
into the same form as eq~\eqref{langevin_equation_with_rouse_ham_matrix}.
Then we have the following Rouse-Ham matrix for this model:
\begin{equation}
 \label{rouse_ham_matrix_two_friction_coefficients}
 \bs{A} =
  \begin{bmatrix}
   4 \phi & -\phi & -\phi & -\phi & -\phi \\
   -1 & 1 & 0 & 0 & 0 \\
   -1 & 0 & 1 & 0 & 0 \\
   -1 & 0 & 0 & 1 & 0 \\
   -1 & 0 & 0 & 0 & 1 
  \end{bmatrix}.
\end{equation}
The Rouse-Ham matrix by eq~\eqref{rouse_ham_matrix_two_friction_coefficients}
is not symmetric, unlike that by eq~\eqref{rouse_ham_matrix}.
We need the left eigenvectors of $\bs{A}$, which correspond to the transpose
of the right eigenvectors of $\bs{A}^{\mathrm{T}}$. Therefore,
we calculate the right eigenvectors of $\bs{A}^{\mathrm{T}}$ in what follows.

The zeroth eigenvalue and eigenvector of $\bs{A}^{\mathrm{T}}$
are $\lambda_{0} = 0$ and  $\bs{v}_{0} = [1 \ \phi \ \phi \ \phi \ \phi]^{\mathrm{T}}$.
This gives the following zeroth normal mode:
\begin{equation}
 \label{tetra_star_normal_mode_0}
  \bm{X}_{0}(t) = \frac{1}{1 + 4 \phi} \left[ \bm{r}_{0}(t) + \phi \sum_{j = 1}^{4} {r}_{j}(t) \right].
\end{equation}
(If $\phi = 1$, eq~\ref{tetra_star_normal_mode_0} reduces to eq~\eqref{zeroth_normal_mode}.
But in this case, $\phi \gg 1$ and thus $\bm{X}_{0}(t) \approx (1/4) \sum_{j = 1}^{4} \bm{r}_{j}(t)$,
which corresponds to the center of mass position of arm beads.)
The fourth eigenvalue and eigenvector of $\bs{A}^{\mathrm{T}}$ are $\lambda_{4} = 1 + 4 \phi$ and
$\bs{v}_{4} = [- 4 \ 1 \ 1 \ 1 \ 1]^{\mathrm{T}}$. This gives the following
fourth normal mode:
\begin{equation}
 \label{tetra_star_normal_mode_5}
 \bm{X}_{4}(t) = \bm{r}_{0}(t) - \frac{1}{4} \sum_{j = 1}^{4} \bm{r}_{j}(t).
\end{equation}
Interestingly, the remaining degenerate eigenmodes are exactly the same as
those for the model with the common friction coefficient: $\lambda_{1} = \lambda_{2} = \lambda_{3} = 1$
and $\bs{v}_{p} = [0 \, s_{p,1} \ s_{p,2} \ s_{p,3} \ s_{p,4}]^{\mathrm{T}}$ ($p = 1, 2,$ and $3$)
with
\begin{equation}
 \begin{bmatrix}
  \bs{s}_{1} & \bs{s}_{2} & \bs{s}_{3}
 \end{bmatrix}
 =
 \begin{bmatrix}
   1 &  0 &  0 \\
  -1 &  1 &  0 \\
   0 & -1 &  1 \\
   0 &  0 & -1
 \end{bmatrix}.
\end{equation}

Zhang et al\cite{Zhang-Tang-Chen-Kwon-Matsumiya-Watanabe-2024a,Zhang-Tang-Chen-Kwon-Matsumiya-Watanabe-2024b} used three span vectors defined as
\begin{equation}
 \Delta \bm{u}_{p}(t) = \sum_{j = 1}^{4} s_{p,j} \bm{r}_{j}(t) \quad (\text{$p = 1, 2,$ and $3$}),
\end{equation}
to analyze the dynamics.
They constructed the dynamic equations for $\lbrace \Delta \bm{u}_{p}(t) \rbrace$ and
explained nonlinear rheological behavior of an end-associative star chain solution.
Although their analyses based on the span vectors seem to be reasonable,
in general, orthogonal normal modes are
preferred than non-orthogonal span vectors.
According to our results in Sec.~3, we can construct the
normal modes in a simple and highly symmetric form for $n = 4$.
From eqs~\eqref{permutation_eigenvectors_4} and \eqref{normal_mode_from_orthogonal_eigenvectors_perm}, we have
\begin{align}
 \label{tetra_star_normal_mode_1}
 \bm{X}_{1}(t) & = \bm{r}_{1}(t) - \bm{r}_{2}(t) + \bm{r}_{3}(t) - \bm{r}_{4}(t), \\
 \label{tetra_star_normal_mode_2}
 \bm{X}_{2}(t) & = \bm{r}_{1}(t) + \bm{r}_{2}(t) + \bm{r}_{3}(t) - \bm{r}_{4}(t), \\
 \label{tetra_star_normal_mode_3}
 \bm{X}_{3}(t) & = \bm{r}_{1}(t) - \bm{r}_{2}(t) - \bm{r}_{3}(t) + \bm{r}_{4}(t).
\end{align}
We expect that model in Refs.~\cite{Zhang-Tang-Chen-Kwon-Matsumiya-Watanabe-2024a} and
\cite{Zhang-Tang-Chen-Kwon-Matsumiya-Watanabe-2024b} would be reformulated in a mathematically
clearer way with the normal modes given by eqs~\eqref{tetra_star_normal_mode_0}, \eqref{tetra_star_normal_mode_5}, and \eqref{tetra_star_normal_mode_1}-\eqref{tetra_star_normal_mode_3}.
(It should be noted that $\lambda_{4} \gg 1$ and the fourth normal mode
relaxes quite rapidly, compared with the first, second, and third normal modes.
Thus it may be ignored in practice, and we can simply assume that $\bm{r}_{0}(t)$ is always fully equilibrated, as in Ref.~\cite{Zhang-Tang-Chen-Kwon-Matsumiya-Watanabe-2024a}.)

\subsection{Gaussian chain model}

In Sec.~2, we employed a highly coarse-grained star polymer model.
To study the relaxation behavior of individual arms, our model is too coarse-grained.
Gaussian-chain-based models would be preferred in some cases.
Here we consider a model in which the ends of $n$ Gaussian chains are
chemically connected to a branch point\cite{Watanabe-Yoshida-Kotaka-1990,Watanabe-Matsumiya-Inoue-2002}.
It would be possible to formulate the Rouse-Ham matrix for 
a star Gaussian chain, but the direct calculations would be complicated.
Instead of direct construction of the Rouse-Ham matrix, we utilize 
some properties of eigenmodes for such a symmetric star Gaussian chain.
The zeroth mode is the same as the coarse-grained model. It simply
represents the center of mass position.

Higher order eigenmodes can be categorized into two types\cite{Watanabe-Yoshida-Kotaka-1990}. One is
so-called the odd mode, where two arms move in an antisymmetric manner.
Another is so-called the even mode, where all the arms move in a symmetric
manner.
In both cases, the deformation of a single arm is essentially
expressed as the eigenmode of a linear Gaussian chain (the Rouse mode).
Odd modes are degenerate, and there are $(n - 1)$ eigenmodes which
have the same eigenvalue. On the other hand,
even modes are not degenerate.
If we concentrate only on the $(n - 1)$ degenerate odd-modes, and
interpret ``a Rouse mode for an arm'' as ``an arm bead'' in our coarse-grained model,
the Gaussian chain model reduces to our coarse-grained model.
Then we can construct the orthogonal normal modes in
the same way as in Sec.~3.

\section{CONCLUSIONS}

We considered how to construct orthogonal normal modes for the degenerate
eigenmodes of the Rouse-Ham type symmetric star chain model.
We showed that we can construct normal modes by combining
Ham's span vectors. For $n = 4$, the permutation-based approach gives
simple and symmetric normal modes. This approach is applicable only for
$n = 4$. For $n = 4, 8, 12, \dots$, the columns of the Hadamard matrix 
can be used to construct the normal modes. The Hadamard matrix exists
only for some limited orders, and thus this method cannot be used if
the Hadamard matrix of order $n$ does not exist.
As a construction method for an arbitrary $n$, we showed the construction based on
the DFT matrix. The span vectors can be interpreted as the difference
operator in one dimensional periodic system, and the columns of the DFT
can be used to construct the normal modes. We expect that the results of this work
can be utilized to analyze various dynamics models based on the Rouse-Ham
model.

\section*{ACKNOWLEDGMENT}
This work was supported by Grant-in-Aid (KAKENHI) for Scientific Research
Grant B No.~JP23H01142.


\end{document}